\documentclass{article}
\usepackage{spconf,amsmath,epsfig, amssymb}
\usepackage{url}
\usepackage{gensymb}
\usepackage[compact]{titlesec}
\usepackage[skip=5pt]{caption}

\title{Ground truth simulation for deep learning classification \\
of mid-resolution VENuS images via unmixing \\
of high-resolution hyperspectral FENIX data}

\twoauthors
 {Ido Faran, Nathan S. Netanyahu}{Eli (Omid) David}
	{Bar-Ilan University\\
	Dept. of Computer Science\\
	Ramat-Gan 5290002, Israel}
 {Maxim Shoshany, Fadi Kizel}{Jisung Geba Chang, Ronit Rud}
	{Technion Israel Institute of Technology\\
	Faculty of Civil and Environmental Engineering\\
	Haifa 3200003, Israel}

\usepackage[overlay,absolute]{textpos}

\begin{document}

\begin{textblock*}{10in}(14mm, 12mm)
{\textbf{Ref:} \emph{IEEE International Geoscience and Remote Sensing Symposium (IGARSS)}, pages 807--810, Yokohama, Japan, July 2019.}
\end{textblock*}

\maketitle

\begin{abstract}
Training a deep neural network for classification constitutes a major problem in remote sensing due to the lack of adequate field data. Acquiring high-resolution ground truth (GT) by human interpretation is both cost-ineffective and inconsistent. We propose, instead, to utilize high-resolution, hyperspectral images for solving this problem, by unmixing these images to obtain reliable GT for training a deep network. Specifically, we simulate GT from high-resolution, hyperspectral FENIX images, and use it for training a \textit{convolutional neural network} (CNN) for pixel-based classification. We show how the model can be transferred successfully to classify new mid-resolution VEN{\boldmath$\mu$}S imagery.

\end{abstract}

\section{Introduction}
\label{sec:intro}
Pixel-based classification of hyperspectral images is a major task in remote sensing, which involves assigning a class label to every pixel of an input image. This task, also known as pixel-wise classification or semantic segmentation, has attracted many studies over the years. Various methods have been proposed for this task. The traditional approach classifies hand-crafted features, using support vector machines \cite{melgani2004classification}, morphological profiles \cite{fauvel2008spectral}, sparse representation \cite{chen2011hyperspectral}, etc. 

However, these methods rely typically on human expertise for tuning them on a specific dataset and they can extract only ``shallow'' features of the original data \cite{bioucas2013hyperspectral}. An alternative approach is to extract useful features directly from the image pixels. 

\textit{Deep learning} (DL) models have proven to be suitable for this kind of problem \cite{zhang2016deep}. Such models are trained on image data sets, and are capable of learning both low-level and high-level feature representations directly from an input image, due to their deep hierarchical architectures. In addition, some DL models can exploit both spectral and spatial features of hyperspectral images, leading to improved classification results.

DL models can be categorized to supervised and unsupervised models. Unsupervised models (e.g., auto-encoders), are trained to extract features from large unlabeled data sets. By restricting the encoder-decoder structure, one can adjust the model to achieve the results of the required task \cite{kemker2017self} \cite{lin2013spectral} \cite{tao2015unsupervised}. Supervised models (e.g., \textit{convolutional neural networks} (CNNs) and deep belief networks) are trained using ground truth (GT) information as expected output of the network. In principle, supervised networks can learn more precise features by exploiting the label-specific information from the training data \cite{hu2015deep} \cite{makantasis2015deep} \cite{yue2015spectral}.

Although most supervised models achieve superior classification, they rely on a considerable amount of GT for training the model. Additionally, there is a limited amount of labeled datasets in the remote sensing community \cite{zhang2016deep}, especially for a new source of information such as a new satellite.

The \textit {Vegetation and Environment monitoring New Micro-Satellite} (VEN{\boldmath$\mu$}S) is a new satellite that was launched in August 2017. It acquires frequent, high-resolution multispectral images of over 100 sites of interest around the world. This enables monitoring of plant growth and their health status, as well as the impact of environmental factors, such as human activities and climate change, on land surfaces of the Earth~\cite{VenusSite}.

Up to this day, there is a relatively small number of images acquired by VEN{\boldmath$\mu$}S and virtually no GT for training a model on this data. Thus, in an attempt to overcome the lack of labeled VEN{\boldmath$\mu$}S data, and in order to start using these images in supervised models, we need to generate GT correlated with the satellite data acquired. 

To avoid the expensive task of obtaining a large number of labeled samples, we propose a novel method for simulating GT from a higher spectral resolution airborne sensor, and using it as initial training data for a CNN model. By applying a state-of-the-art spectral unmixing algorithm~\cite{kizel2017stepwise} to the above airborne data, and adapting the acquired images to the spatial and spectral resolutions of VEN{\boldmath$\mu$}S, we can train a CNN to classify VEN{\boldmath$\mu$}S images to several predefined endmembers (EMs). This approach may help provide initial classification of incoming VEN{\boldmath$\mu$}S images, without any GT, as part of a more comprehensive effort of processing time-series VEN{\boldmath$\mu$}S data on a continuous basis.

The paper's contributions are as follows: (1) Introduction of a  novel GT simulation for training a DL-classification model without manual labeling, (2) presentation for the first time of classification results for the recently launched VEN{\boldmath$\mu$}S satellite over a Mediterranean region that is of much interest (as far as climate change is concerned), i.e., the results can serve as a baseline for comparison with further methods, and (3) providing simulated data that will allow us to train more sophisticated models (such as fully convolutional neural network~\cite{long2015fully}) or explore more complex tasks, such as spectral unmixing using neural networks, for further processing of VEN{\boldmath$\mu$}S data.

\section{Background}
\label{sec:format}

\subsection{FENIX}

\begin{figure}[tb]
  \centering
  \centerline{\epsfig{figure=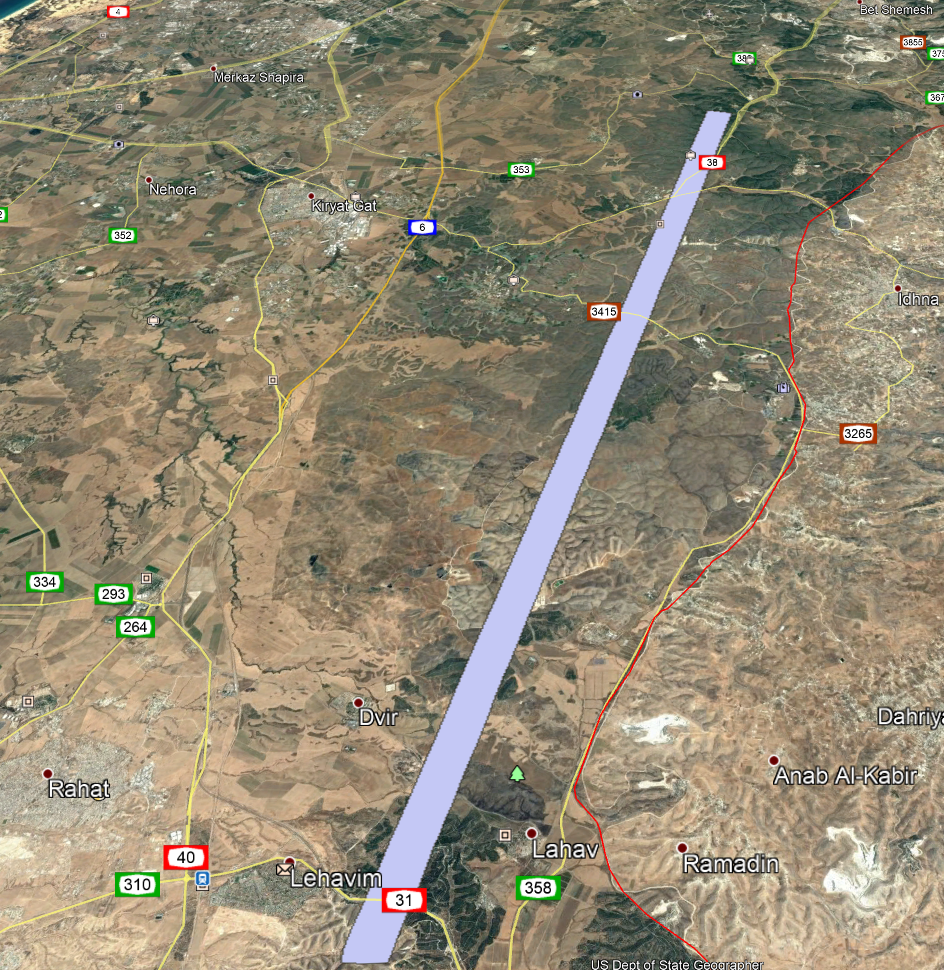,width=3.0cm}}
  \caption{FENIX flight strip over Israel}
\label{fig:FENIXStrip}
\end{figure}

A FENIX airborne scan took place on April 04, 2017 under clear sky conditions, along a transect from the Beit-Guvrin area (representing a semi-arid region with a rainfall rate of 450[mm/year]) to Lehavim (representing a desert fringe zone with 250[mm/year] rainfall). The scan was carried out over a 35-[km] long strip with a swath of 1.5[km] (Figure~\ref{fig:FENIXStrip}). A SPECIM AisaFENIX 1K airborne system was mounted on a Cessna 172 airplane flying at an altitude of 1,828[m] above sea level over a topographic area with an average height of 250[m]. The system consisted of VNIR and SWIR instruments, yielding a ground sampling distance of 1[m], and having wavelength ranges of 380-970[nm] (at 4.5[nm] spectral resolution) and 970-2500[nm] (at 12[nm] spectral resolution), respectively. The spectral bands were resampled into 41 bands of 5[nm] width in the wavelength range of 400-2400[nm].

\subsection{VEN{\boldmath$\mu$}S}
The VEN{\boldmath$\mu$}S satellite carries a super-spectral camera characterized by 12 narrow spectral bands ranging from 415[nm] to 910[nm]. The radiometric resolution for all bands is 10 bits and the spatial resolution is 5[m]. The spectral resolution at the Vis-NIR range is 40[nm], and 16[nm] and 20[nm], respectively, for the red edge and water vapor bands. Each experimental site is of size $27 \times 27$[km], with a 2-day revisit time.

\subsection{Spectral Unmixing}
\noindent Given a spectral image and the spectra of a set of distinct EM materials,
the spectral unmixing process allows for extracting quantitative subpixel information by estimating the abundance fraction of each EM, in each pixel. Assuming a \textit{linear mixture model} (LMM), we write the spectral signature of each pixel as follows:\[\mathbf{m=Ef+n}\] 
where $\mathbf{m}=[m_{1} ,\, ...,\, m_{\lambda }]^{T} $ is a signature of mixed pixel, $\lambda $ is the number of spectral bands, $\mathbf{E}\in {\rm R}^{\lambda \times d} $ is the matrix of $d$ EMs, $\mathbf{f}\in {\rm R}^{d\times 1} $ is a vector of the corresponding fractions, and $\mathbf{n}\in {\rm R}^{\lambda \times 1} $ represents the system noise and assumed to be Gaussian with zero mean. Requiring a fully constrained solution, the unmixing problem is solved subject to two constraints: $f_{i} \ge 0$ for $i=1,\, \ldots ,\, d$, and $\mathbf{f^{{\rm T} } 1}\le 1$, where $\mathbf{1}\in {\rm R}^{d\times 1} $ is a vector of ones. In our case, we use the \textit{vectorized code projected gradient descent unmixing} (VPGDU) method~\cite{kizel2017stepwise}. VPGDU combines the \textit{projected gradient descent} (PGD) and an exact line search strategy to optimize an objective function that is based on \textit{spectral angle mapper} (SAM). 

\subsection{Convolutional Neural Network (CNN)}
CNNs have shown excellent performance in various visual perception tasks, such as object detection, object classification, semantic segmentation, etc., by exploiting the local connectivity between adjacent pixels. 
Recently, CNNs have also been used successfully for classification of hyperspectral images; see, e.g., Hu et al.~\cite{hu2015deep}, Makantasis at el.~\cite{makantasis2015deep}, and Yue et al.~\cite{yue2015spectral}.

\section{PROPOSED METHOD}
\label{sec:pagestyle}

\subsection{Overview}
\begin{figure}[tb]
  \centering
  \centerline{\epsfig{figure=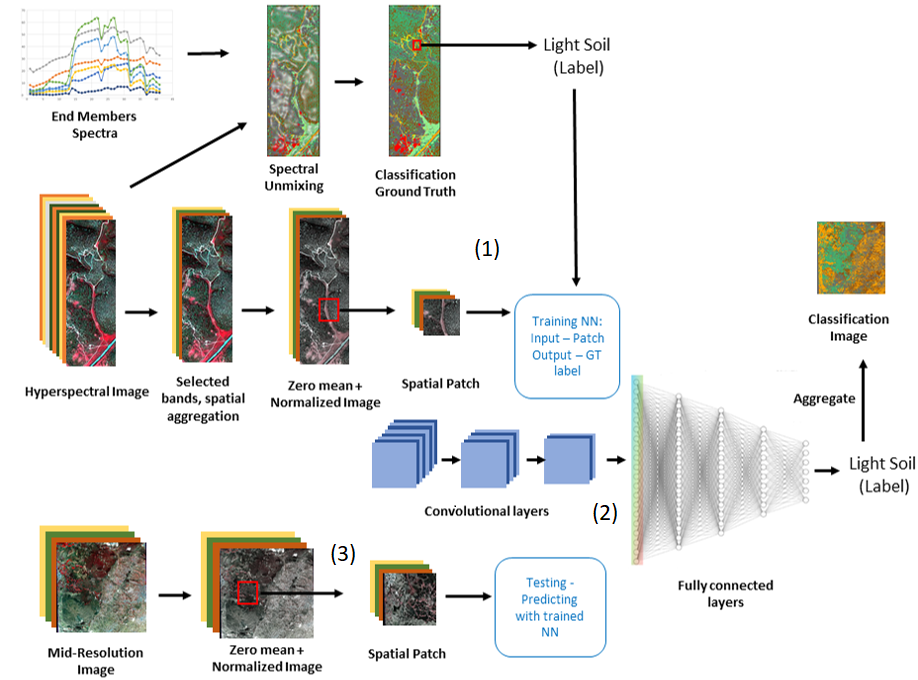,width=8.0cm}}
  \caption{Architecture of proposed method: CNN trained using simulated GT from hyperspectral image, and then used for classification of mid-resolution image.}
\label{fig:architecture}
\end{figure}

Figure~\ref{fig:architecture} illustrates the framework of the proposed method. It consists of three parts:
(1) Ground truth simulation, (2) training a CNN model, and (3) evaluating its classification on a real VEN{\boldmath$\mu$}S image. In the first part, a spectral unmixing algorithm is executed on higher-resolution images using predefined labels and their estimated abundance vectors in order to extract fraction vectors. The original images are then aggregated and adjusted to match VEN{\boldmath$\mu$}S's spatial and spectral resolutions. 
In the second step, we use spatial patches around each labeled pixel to train a deep CNN. Finally, we apply the trained network to a calibrated VEN{\boldmath$\mu$}S L1 image to obtain its classification map. The proposed method is described below in detail.

\subsection{Ground Truth Simulation}
\begin{figure}[tb]
  \centering
  \centerline{\epsfig{figure=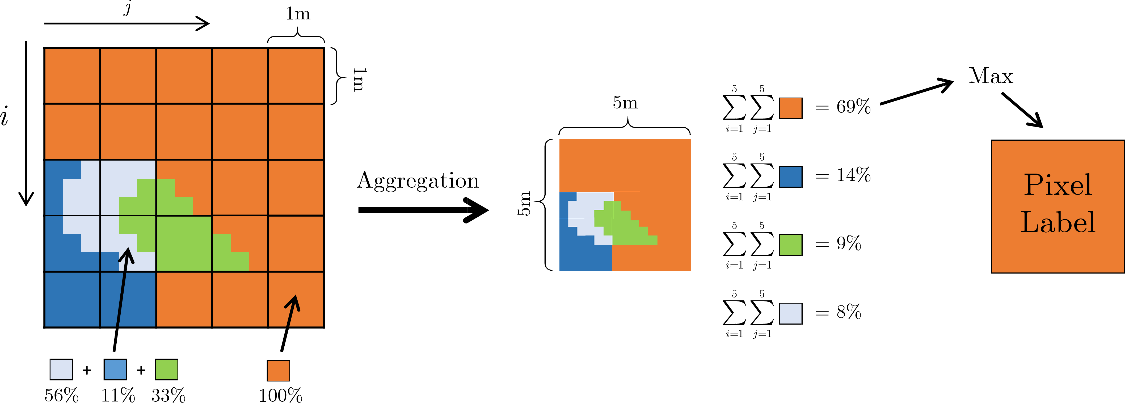,width=8.0cm}}
  \caption{Illustration of synthesized GT pixel labels based on unmixing results.}
\label{fig:gtCreation}
\end{figure}

We simulate plausible GT for VEN{\boldmath$\mu$}S by converting a given FENIX image (at 1[m] resolution). Non-contiguous regions of $5 \times 5$ pixels are aggregated to a single pixel (at 5[m] resolution), and only the bands matching VEN{\boldmath$\mu$}S's spectral resolution are selected. To synthesize the GT labels (Figure \ref{fig:gtCreation}), we first generate fraction maps of the high-resolution FENIX image (by applying VPGDU with respect to the seven EMs selected). The label assigned to a given pixel of the simulated VEN{\boldmath$\mu$}S image is the EM for which the aggregated fractions (over the corresponding $5 \times 5$ region in the FENIX image) is the greatest.

\subsection{Training Neural Networks}
The simulated input images are split to small patches \cite{yang2017learning} \cite{chen2014deep}. Each patch contains a spatially correlated area around a specific pixel and its label as explained above. This allows for creating a large amount of label samples for training. 

We examined several DL models for the classification task; the best results were achieved for a deep CNN model. The proposed network receives an $n \times m \times b$ input matrix (where $n$, $m$ are the patch dimensions, and $b$ is the number of spectral bands). It consists of $L$ convolution layers with a decreasing amount of filters per layer. Because of the relatively small input size, max-pooling layers are not necessary to simplify the dimensions of the model. Finally, the output of the last convolution layer is flattened and fed into a number of fully connected layers. The softmax activation function is applied to the last layer's output for creating classification fractions. The output label of the center pixel of the patch is determined by the class of the greatest fraction.

In addition, due to the unbalanced amount of samples per label, data augmentation and denoising layers are added to prevent the network from over-fitting to the most frequent labels. Adding Gaussian noise, random rotations and image mirroring are used for creating balanced label counts per sample. Batch normalization~\cite{ioffe2015batch} and dropout~\cite{srivastava2014dropout} layers are also added between hidden layers to insert noise into the training process.

\subsection{Evaluating on New Dataset}
The classification due to the trained network can be ``transferred'' to new images acquired over similar geographical regions with the same EMs. This is applied to VEN{\boldmath$\mu$}S images at the L1 level (that were geographically calibrated to clean background noises while preserving spatial resolution). Image patches (from the real VEN{\boldmath$\mu$}S image) are then fed into the trained network as before to obtain a classification label for their center pixels.

\section{EXPERIMENTAL RESULTS}
\label{sec:typestyle}

\subsection{Datasets}
The suggested procedure has been tested on simulated and real VEN{\boldmath$\mu$}S images for quantitative/visual assessment.

The simulated training data was acquired from the FENIX dataset by taking the the six most relevant areas, with an average picture size of $400 \times 100$ pixels.

The VEN{\boldmath$\mu$}S test data of size $6829 \times 7824$ pixels, with a spatial resolution of 5[m] and 11 spectral bands \footnote{band 6 is removed, as it is a duplication of band 5 for image quality}, was acquired over the S02 polygon of Israel~\cite{VenusIsraelWebSite} on June 15, 2018. We worked with atmospherically corrected L1 products to maintain the original spatial resolution.

The following seven EMs that match the common land composition in this area were selected: Brown Soil, Light Soil, Rock, Tall Tree/Shrub, Dwarf Shrub, Herbaceous, and Dense Shrub/Burned Area.

\subsection{Parameters and Details}
To obtain reliable results, we conducted a 6-fold cross validation; each time one of the images was left out for testing, and the rest were used for training and validation. Specifically, all of the pixels of the latter images were randomly shuffled, and each time 90\% of these pixels were used for training and the remaining 10\% of the pixels for validation. Also, we normalized the images, as part of preprocessing, by standardizing the values of each spectral band to have zero mean and a standard deviation of 1.0.

After examining several patch configurations, we selected $5 \times 5$ patches (i.e., $25[m] \times 25[m]$ regions) around each pixel, labeled according to the center pixel of the patch. To train a balanced model with a similar count of samples per label, data augmentation was applied via horizontal/vertical flips, rotations by 90\degree, 180\degree, and 270\degree, and addition of Gaussian noise with zero mean and 0.1 standard deviation. During each epoch, 30,000 samples per label (for a total of 210,000 samples in each epoch) were created using a combination of the above techniques.

The full network architecture is shown in Figure~\ref{fig:architecture}. For the CNN model, we used 4 layers of $3 \times 3$ convolution filters, with a different amount of filters per layer, i.e., 64, 64, 32, and 16 filters, respectively. Batch normalization layers are used (before applying a ReLU activation function), as well as dropout layers with a rate of 25\%. The CNN is followed by 3 fully connected layers with an output size of 7 neurons. (The hidden layers are activated using the ReLU function, while the output layer uses softmax activation.) 

The following hyperparameters were arrived at after various tuning attempts: Batch size = 64, cross entropy loss function, and Adam 
optimizer with a learning rate of 0.001. All weights were randomly initialized. The deep neural network was implemented using the Python programming language with
TensorFlow as the DL framework. The network was trained over 200 epochs using backpropagation on a PC equipped with Intel Core I7 and Nvidia GeForce GTX 1080 Ti GPU.

\begin{figure}[t]
\begin{minipage}[b]{.3\linewidth}
  \centering
  \centerline{\epsfig{figure=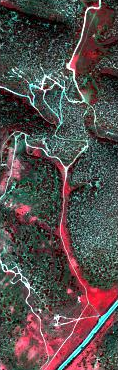,width=2.0cm}}
  \centerline{(a) False color composite}\medskip
\end{minipage}
\hfill
\begin{minipage}[b]{.3\linewidth}
  \centering
  \centerline{\epsfig{figure=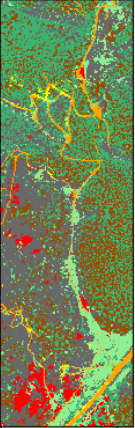,width=2.0cm}}
  \centerline{(b) Ground Truth}\medskip
\end{minipage}
\hfill
\begin{minipage}[b]{0.3\linewidth}
  \centering
  \centerline{\epsfig{figure=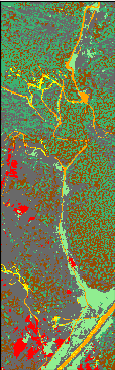,width=2.0cm}}
  \centerline{(c) CNN classification}\medskip
\end{minipage}
\begin{minipage}[b]{1.0\linewidth}
  \centering
  \centerline{\epsfig{figure=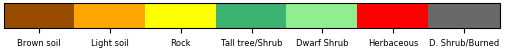,width=9cm}}
\end{minipage}
\caption{Visualization of the results obtained over the Avisure2 area.}
\label{fig:FENIXResultsVis}
\end{figure}

\begin{figure*}[t]
\begin{minipage}[b]{.3\linewidth}
  \centering
  \centerline{\epsfig{figure=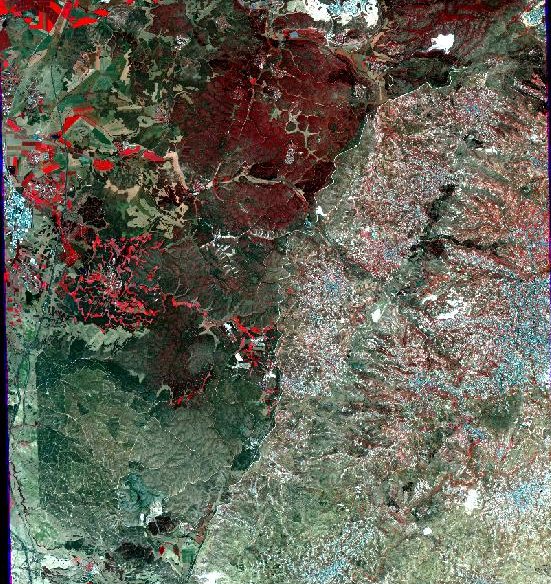,width=5cm}}
  \centerline{(a) False color composite}\medskip
\end{minipage}
\hfill
\begin{minipage}[b]{.3\linewidth}
  \centering
  \centerline{\epsfig{figure=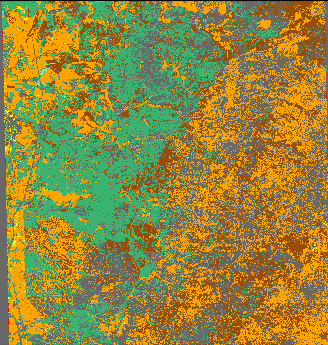,width=5cm}}
  \centerline{(b) CNN classification}\medskip
\end{minipage}
\hfill
\begin{minipage}[b]{0.3\linewidth}
  \centering
  \centerline{\epsfig{figure=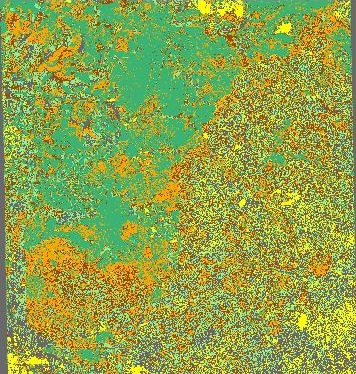,width=5cm}}
  \centerline{(c) ISODATA output}\medskip
\end{minipage}
\caption{Visualization of the results on a real VEN{\boldmath$\mu$}S image.}
\label{fig:VenusResultsVis}
\end{figure*}

\subsection{Results}
Table~\ref{tab:tableAccuracy} reports the classification results obtains by the proposed network on the 6 test images. The average accuracy obtained was $\sim72.6\%$.
Although the quantitative results are not extremely high, visual assessment reveals a notable similarity between GT and the classification maps for large parts of the simulated and real VEN{\boldmath$\mu$}S images (Figures~\ref{fig:FENIXResultsVis} and~\ref{fig:VenusResultsVis}, respectively). This should attest to the good promise of our baseline method for further classification of new VEN{\boldmath$\mu$}S images.

\begin{table}
\begin{center}\
 \begin{tabular}{| c || c | c | } 
 \hline
 Testing Image & Validation Accuracy & Test Accuracy \\ [0.5ex] 
 \hline\hline
 Amazya1 & 80.67\% & 69.40\% \\ 
 \hline
 Avisure1 & 82.65\% & 69.97\% \\
 \hline
 Avisure2 & 80.61\% & 75.08\% \\
 \hline
 Between1 & 80.48\% & 70.96\% \\
 \hline
 Between2 & 82.49\% & 73.27\% \\
 \hline
 Lehavim1 & 80.80\% & 76.71\% \\
 \hline
 \hline
 \textbf{Overall}  & \textbf{81.29\%} & \textbf{72.56\%} \\
 \hline
\end{tabular}
\end{center}
\caption{\label{tab:tableAccuracy}Classification results on FENIX test images.}
\end{table}

\section{CONCLUSION}
\label{sec:majhead}
We proposed a novel method for GT simulation of mid-resolution data by applying unmixing to high-resolution hyperspectral images. This allows to overcome a fundamental problem in remote sensing, i.e., a severe lack of labeled data. The simulated data was used for initial training of a CNN for pixel-based classification, as part of an ongoing project of temporally evolving CNNs for the analysis of the ecological mapping of Mediterranean environments using VEN{\boldmath$\mu$}S images.

\bibliographystyle{IEEEbib}
\bibliography{venus-images}

\end{document}